\documentclass[10pt]{article}
\usepackage[margin=1.2in]{geometry}
\usepackage{authblk}
\usepackage{booktabs}
\usepackage{amsmath}
\usepackage{url}

\usepackage{breakurl}
\usepackage{graphicx}
\usepackage[numbers]{natbib}

\newcommand{\ra}[1]{\renewcommand{\arraystretch}{#1}}

\title{Social capital predicts corruption risk in towns}

\author[a,\thanks{The authors thank Bernie Hogan, Ralph Schroeder, David Deritei, Mih\'aly Fazekas, Marc Sarazin, Cohen Simpson, Bharath Ganesh, and participants of seminars at Oxford and MTA for valuable insights. The authors are grateful to J\'anos T\"or\"ok for assistance with the iWiW data, and to \'Agnes Czibik and Mih\'aly Fazekas for assistance with the public contracting data. TY was partially supported by the Alan Turing Institute under the EPSRC grant no. EP/N510129/1.}]{Johannes Wachs}
\author[b,c]{Taha Yasseri} 
\author[d,e]{Bal\'azs Lengyel}
\author[a,f]{J\'anos Kert\'esz}

\affil[a]{Department of Network and Data Science, Central European University, H-1051 Budapest, Hungary}
\affil[b]{Oxford Internet Institute, University of Oxford, 1 St Giles, Oxford OX1 3JS, UK}
\affil[c]{Alan Turing Institute, 96 Euston Road, London NW1 2DB, UK}
\affil[d]{Agglomeration and Social Networks Lend\"ulet Research Group, Hungarian Academy of Sciences, H-1097 Budapest, Hungary}
\affil[e]{International Business School Budapest, H-1037 Budapest, Hungary}
\affil[f]{Institute of Physics, Budapest University of Technology and Economics, H-1111 Budapest, Hungary}

\begin{document}

\maketitle
\begin{abstract}
Corruption is a social plague: gains accrue to small groups, while its costs are borne by everyone. Significant variation in its level between and within countries suggests a relationship between social structure and the prevalence of corruption, yet, large scale empirical studies thereof have been missing due to lack of data. In this paper we relate the structural characteristics of social capital of towns with corruption in their local governments. Using datasets from Hungary, we quantify corruption risk by suppressed competition and lack of transparency in the town's awarded public contracts. We characterize social capital using social network data from a popular online platform. Controlling for social, economic, and political factors, we find that settlements with fragmented social networks, indicating an excess of \textit{bonding social capital} have higher corruption risk and towns with more diverse external connectivity, suggesting a surplus of \textit{bridging social capital} are less exposed to corruption. We interpret fragmentation as fostering in-group favoritism and conformity, which increase corruption, while diversity facilitates impartiality in public life and stifles corruption.

\end{abstract}

\section*{Introduction}
Corruption is widely recognized to affect adversely social and economic outcomes of societies~\cite{mauro2004persistence}, yet it is difficult to fight~\cite{mungiu2013controlling}. Though education and income seem to decrease corruption~\cite{glaeser2006corruption}, it persists even under highly developed, democratic conditions, however, showing significant regional differences~\cite{charron2014regional}. Researchers often relate corruption to social aspects of society such as segregation~\cite{alesina2011segregation}, interpersonal trust~\cite{rothstein2005all}, civic-mindedness~\cite{guiso2011civic}, and community engagement~\cite{putnam2001bowling}. These approaches build on the insight that corruption is a collective outcome of a community shaped by the interactions among individuals~\cite{persson2013anticorruption}, suggesting that differences in the social capital, especially in the network structure may help explain the persistence of corruption and the observed differences in its levels.

The concept of social capital, or the "connections among individuals – social networks and the norms of reciprocity and trustworthiness that arise from them"~\cite{putnam2001bowling}, is usually applied to understand behavior of individuals~\cite{portes1998social}. Yet city or country-level  aggregations have also proven useful~\cite{putnam2001bowling}, for example, in studying economic development and prosperity~\cite{westlund2010social}. As a communal quantity, social capital is a sort of public good embedded in a social network~\cite{lin2002social} of a settlement. It is therefore natural to expect that the structure of social capital at settlement level has considerable impact on the scale of corruption in that community. Despite significant interest in the network aspects of corruption~\cite{gambetta1996sicilian} and recent experimental evidence that corruption has collaborative roots~\cite{weisel2015collaborative}, less is known about how the patterns of connectivity of a whole society influences the general level of corruption in its government.

Previous studies of relating social capital and corruption~\cite{harris2007,bjornskov2003corruption} have been constrained by two empirical challenges: the difficulty of measuring corruption and the lack of data on network structure at the settlement level. Corruption is one of the most hidden type of crimes therefore it is difficult to estimate its extent in general. Many studies measure corruption using national or regional surveys~\cite{mungiu2013controlling} and suffer from the subjectivity of corruption perceptions~\cite{torsello2016anthropology}. Other studies use data on the frequency of investigations and convictions of politicians~\cite{glaeser2006corruption}, in which a source of bias may be that in places where corruption is prevalent the judiciary is more likely to be corrupt itself. Recent efforts to clean and standardize large datasets on public procurement~\cite{fazekas2017uncovering} have been very helpful in this context as their study can lead to new, more objective indicators of corruption risk.

In the absence of direct network data, researchers often quantify social capital using proxies such as rates of voting, donating blood, and volunteering~\cite{guiso2011civic}. As these rates are themselves related to the underlying social networks, they indicate rather the relevance of social capital instead of explaining the causes of corruption in terms of network structure. Mapping out the social capital at the level of settlements using traditional tools is a formidable task. Fortunately, recent developments in information-communication technologies and their increasing popularity enable to access large data sets containing relevant information in this respect. For example, data from online social networks and cellphone records have been used to relate connectivity and socio-economic outcomes~\cite{eagle2010network,bokanyi2016race,mamei2018social,norbutas2018network,bailey2018social}.

In this paper we propose to characterize the level of corruption risk in settlements in terms of their social capital using two sources of micro-level data from Hungary. We quantify the structural characteristics of towns' social capital using complete data from ``iWiW'', a now defunct online social network once used by approximately 40\% of the adult Hungarian population~\cite{lengyel2015geographies}. We measure corruption risk using administrative data on public procurement contracts~\cite{fazekas2016objective}. 

Public procurement contracts constitute a major channel of public funds to private hands and are highly vulnerable to corruption~\cite{fazekas2017uncovering}. Recently, a set of corruption risk indicators have been derived from public contract data, for example, if the contract attracted a single bidder only. These contract-based corruption risk measures are significantly related to other proxies of corruption. When aggregated at the regional and national levels, they have been shown to correlate significantly with corruption perception surveys~\cite{fazekas2017uncovering} and quality of government indicators~\cite{charron2017careers}. In the Hungarian case, we find that a group of settlements involved in a recent corruption scandal~\cite{elios} have significantly higher corruption risk in their contracts.

Putnam distinguishes between two structural categories of social capital: \textit{bonding} and \textit{bridging} social capital~\cite{putnam2001bowling} and we expect that these have different impacts on corruption risk. Bonding social capital is based on the phenomenon of closure in a social network, describing the extent to which people form dense, homogeneous groups. Such groups have benefits: members share high levels of trust and can count on each other in times of crisis. They can also be sure that members who defy the norms of the group will be censured. The homogeneity of such tight-knit groups is often based on ethnicity, religion, or class~\cite{coleman1988social} indicating possible drawbacks: homogeneity facilitates conformity and implies exclusion of outsider groups. Solidarity can reach the extent that members of the group will protect each other even if \textit{norms outside the group} are broken, in some cases even if crimes are committed.  Sophisticated criminal organizations like the Mafia, members of which may regularly be faced with great incentives to ``flip'', rely on bonding rituals, ethnic homogeneity, and family ties to enforce solidarity and in-group trust~\cite{erickson1981secret,gambetta1996sicilian}. The negative effects of excessive bonding social capital on society are not limited to crime and corruption. Entrepreneurs embedded in dense networks are disadvantaged because of pressure to employ under-qualified relatives~\cite{geertz1963peddlers}, while ethnically homogeneous groups of  traders are more likely to mis-price financial assets in the same direction and create bubbles~\cite{levine2014ethnic}.

Bridging social capital, on the other hand, refers to the connections in a community between people from different social groups. Such ties are valuable for their ability to convey novel information~\cite{granovetter1977strength} and exposure to diverse perspectives, though they do not serve as strong conduits of trust. Previous work shows, for instance, that immigrants in the Netherlands with bridging connections outside their ethnic group have significantly higher incomes and employment rates~\cite{lancee2010economic}. But bridging social capital is not only thought to be useful for the resources it allocates. Using an agent-based model, Macy and Skvoretz showed how trust emerged among densely connected neighbors and diffused in a social network via weak ties~\cite{macy1998evolution}, implying that low bridging social capital restricted trust to within-group interactions. Indeed, empirical evidence showed that ethnic groups in diverse communities with more bridging social capital evaluate each other more positively~\cite{laurence2009effect}.

We pose two hypotheses relating bonding and bridging social capital to local corruption risk. The first (H1) is that excess bonding social capital in a town enables corruption in its government by fostering norms of in-group favoritism thus it increases corruption risk. The second (H2) is that surplus bridging capital, including connections to other towns, suppresses corruption by fostering impersonal and universalistic norms, thus it decreases corruption risk. The two hypotheses mirror, quoting Portes, ``Durkheim's distinction between mechanical solidarity, based on social homogeneity and tight personal bonds, and organic solidarity, based on role differentiation, impersonal norms, and an extensive division of labor.''~\cite{portes2014downsides}. Where mechanical solidarity or bonding social capital dominate organic solidarity or bridging social capital, universalistic norms under which public markets are thought to function best are unsustainable. These hypotheses suggest why corruption is so difficult to fight: it is embedded in the social network of a place. 

Previous work using survey data is in accord with our hypotheses. Harris finds a significant positive relationship between excess bonding social capital, measured using surveys, and corruption across over 200 countries~\cite{harris2007}. In a comparative study of the 50 US states, Knack finds that residents in states with higher census response and volunteering rate their governments' performances more highly~\cite{knack2002social}. He finds no such effect for rates of membership in social clubs, a more exclusive form of socialization than volunteering. Paccagnella and Sestito find that in regions with high electoral turnout and blood donation rates, Italian schoolchildren cheat less frequently on standardized tests. In schools with greater ethnic homogeneity and with hometown teachers, cheating is more frequent~\cite{paccagnella2014school}. These case studies and indirect evidences give some support the above hypotheses, however, there is need for studies based on more direct data. 

We find significant evidence for our hypotheses using multivariate regression models to relate corruption risk and structural aspects of social capital. Hungarian settlements with fragmented social networks, which we interpret as evidence of excess bonding social capital, have higher corruption risk in their public procurement contracts. On the other hand, if the typical resident of a settlement has more diverse connections especially over the boundaries of their own towns, then the local corruption risk is lower. These results hold controlling for several potential confounders including economic prosperity, education, demographics, and political competitiveness.

\section*{Empirical Setting and Methods}

\subsection*{Public contracting}
In OECD economies procurement counts typically between 10 to 20\% of GDP~\cite{oecdprocurement} covering everything from school lunches to hospital beds and highway construction. The complexity of the contracts and the relative inelasticity of the government's demand for goods make them a prime target for corruption~\cite{pwc2013}.

Contracts are supposed to be awarded using impartial market mechanisms~\cite{weber1978economy}: open and fair competition for a contract is considered the best way to insure that the government makes purchases of good quality at the lowest cost. Usually, an issuer of a contract publishes a call for bids from the private sector, setting a deadline for submissions leaving enough time for broad participation. Companies submit sealed offers, including a price. The company offering to provide the good or service for the lowest price, meeting the standards set in the call for bids, wins the contract.

\subsubsection*{Measuring settlement corruption risk in contracting}
Corruption in public contracting typically involves the restriction of competition. If corrupt bureaucrats wish to award a contract to a favored firm, they must somehow exclude other firms from participating in the competition for the contract. We quantify this phenomenon at the contract level by tracking the presence of elementary corruption indicators, signals we can extract from metadata suggesting that competition may have been curbed~\cite{fazekas2017uncovering}. These quantitative indicators, deduced~\cite{fazekas2016objective} from qualitative work on corruption in public contracting, are the fingerprints of techniques used to steer contracts towards preferred firms. We consider eight such elementary indicators, defined in Table~\ref{table:cri}.

\begin{table}[t]
\ra{1.6}
\begin{tabular}{p{0.3\textwidth}cp{0.5\textwidth}}
\toprule
Indicator and Symbol & Values& Indicator Definition\\
\midrule
Single bidder\newline$C_{singlebid}$ & $\{0,1\}$ & 1 if a single firm submits an offer.\\[.2cm]
Closed procedure\newline$C_{closedproc}$  & $\{0,1\}$ & 1 if the contract was awarded directly to a firm or by invite-only competition. \\[.2cm]
No call for bids\newline$C_{nocall}$  & $\{0,1\}$ & 1 if no call for bids was published in the official procurement journal. \\[.2cm]
Long eligibility criteria\newline$C_{eligcrit}$   & $\{0,1\}$ & 1 if the length in characters of the firm eligibility criteria is above the market average\footnote{We define markets by contract Common Procurement Vocabulary (CPV) codes~\cite{cpvreport}}. \\[.2cm]
Extreme decision period\newline$C_{decidetime}$ & $\{0,1\}$& 1 if the award was made within 5 days of the deadline or more than 100 days following.\\[.2cm]
Short time to submit bids\newline$C_{bidtime}$ & $\{0,.5,1\}$ & 1 if the number of days between the call and submission deadline is less than 5, 0.5 if between 5 and 15.\\[.2cm]
Non-price criteria\newline$C_{nonprice}$ &$\{0,1\}$ & 1 if non-price criteria are used to evaluate bids.\\[.2cm]
Call for bids modified\newline$C_{callmod}$  & $\{0,1\}$& 1 if the call for bids was modified.\\[.2cm]
\bottomrule
\end{tabular}
\caption{Elementary indicators of public contract corruption risk.}\label{table:cri}
\end{table}

\noindent From these eight elementary indicators we define two measures of contract corruption risk:

\textit{Closed procedure or single bidding ($C_{csb}$):} Did the contract attract only a single bid \textbf{or} was the contract awarded by some procedure besides an open call for bids, for example by direct negotiation with a firm or by an invitation-only auction? In terms of the indicators defined above:
$$C_{csb}=max(C_{singlebid},C_{closedproc})$$

\textit{Corruption Risk Index ($CRI$):} Following~\cite{fazekas2016objective}, we average all eight elementary indicators defined in Table~\ref{table:cri} for each contract:\\

\noindent $$CRI = \frac{1}{8} \left(C_{singlebid}+C_{closedproc}+C_{nocall}+C_{eligcrit}+C_{bidtime}+C_{nonprice}+C_{callmod}+C_{decidetime}\right) $$

Such indicator-based measures of corruption risk have been related to traditional measures of corruption at the regional and national levels. Among EU countries, similar indicators are highly correlated with both the World Bank's Control of Corruption rankings and Transparency International's Corruption Perceptions Index~\cite{fazekas2017uncovering}. They also predict cost overruns and price inflation in European infrastructure projects~\cite{fazekas2018extent}. At the micro-level, public bodies issuing high corruption risk contracts are significantly more likely to award contracts to new companies after a change in government~\cite{fazekas2017networks}. Finally, evidence from the US suggests that firms making campaign contributions are awarded contracts with higher corruption risk~\cite{fazekas2018institutional}.

\subsubsection*{Local Government Contracting Data}
We examine 20,524 municipal government contracts from the period of 2006-2014 issued by Hungarian settlements awarding at least five contracts a year on average. There are 169 settlements in Hungary meeting this criterion, excluding Budapest\footnote{Budapest, the capital of Hungary, is an order of magnitude larger than the second largest settlement and has a unique governance structure.}. We average both corruption risk indicators over all contracts issued by a town to arrive at its rate of issuing closed procedure or single-bid contracts and its average $CRI$.

\begin{figure*}
\includegraphics[width=\textwidth]{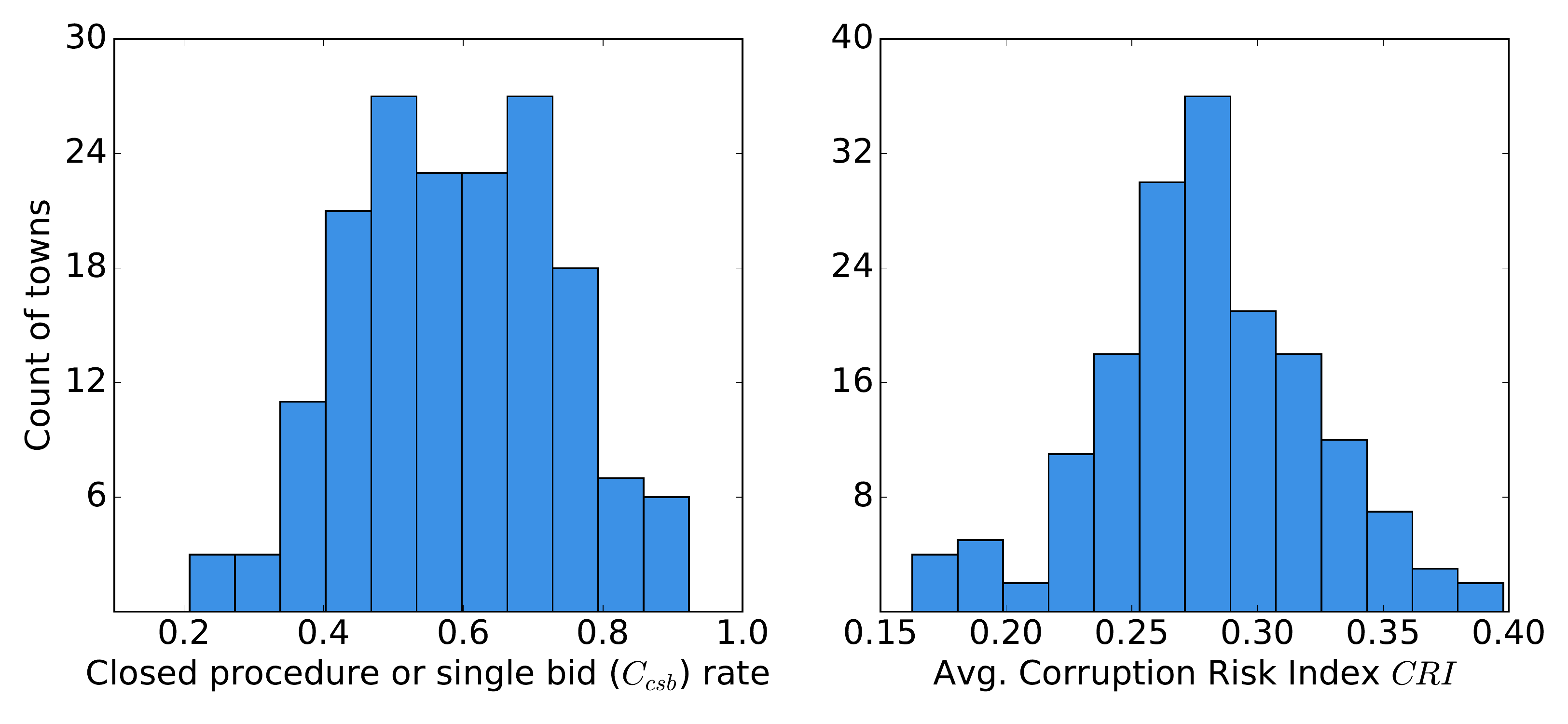}
\caption{Distributions of average contract corruption risk indicators across Hungarian settlements.}
\label{fig:crisk_dists}
\end{figure*}

We plot the distributions of the settlement corruption risk scores in Figure~\ref{fig:crisk_dists}. We note that there is substantial variation across settlements: some award over 90\% of their contracts either via a closed procedure or to a single bidding supplier, while others do so less than 25\% of the time.

As a test of the validity of our settlement-level measures of corruption risk, we check them against a near-ground truth case of corruption. In 2018, OLAF, the European anti-fraud agency reported that 35 Hungarian local government public lighting contracts awarded between 2010 and 2014 contained ``serious irregularities''~\cite{elios,norman_komuves_2018}. Elios, the company winning these contracts, was owned at that time by the son-in-law of the Hungarian Prime Minister. The contracts are considered to be overpriced and the Hungarian government was appealed for initiating an investigation, which has already started.

\begin{figure*}
\includegraphics[width=\textwidth]{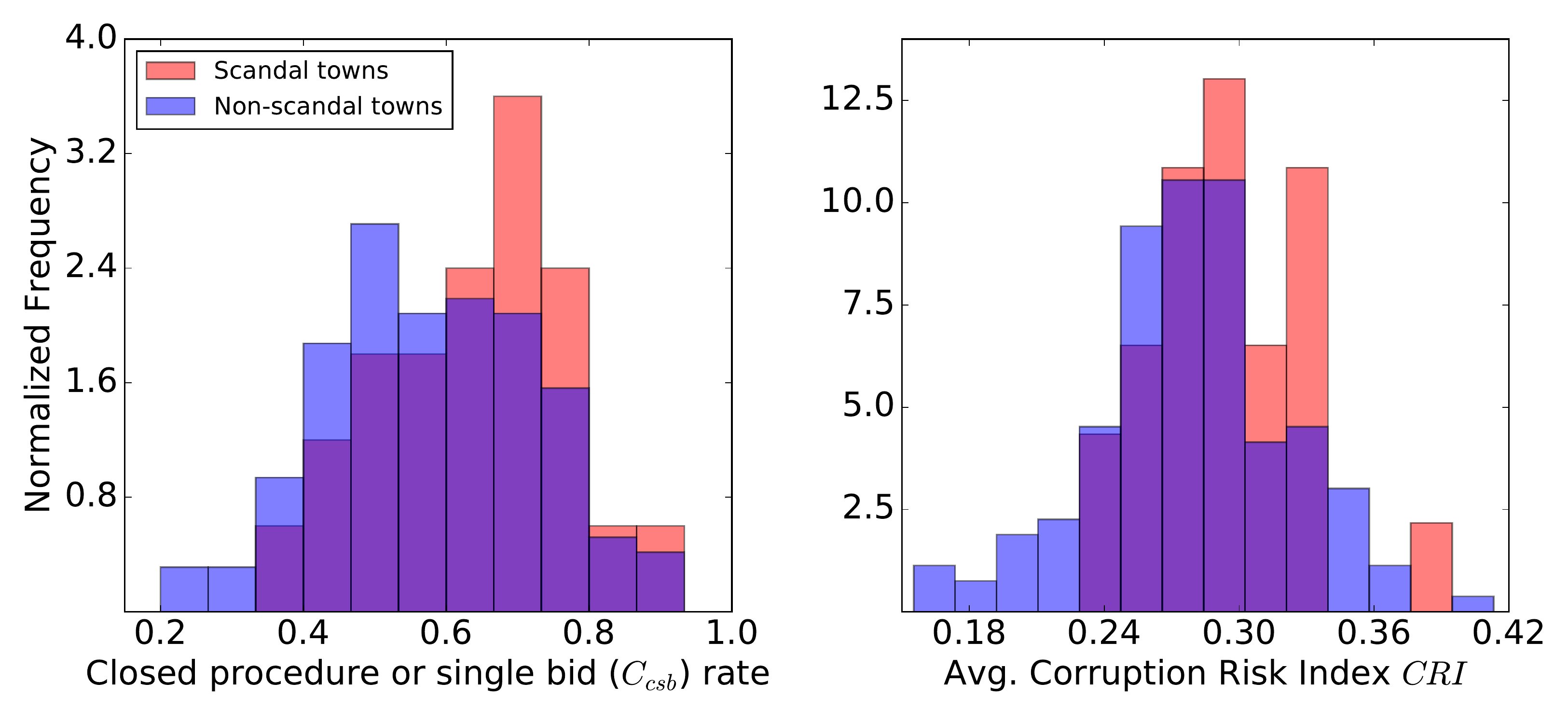}
\caption{Distributions of average contract corruption risk indicators for settlements involved in the Elios scandal compared with all other settlements. Settlements involved in the scandal have significantly higher average corruption risk in their contracting than their counterparts.}
\label{fig:elios_dist}
\end{figure*}

These cases provide a useful test of our corruption risk indicators. There is compelling evidence that settlements implicated in the scandal have, at least once, rigged a public procurement contract to favor a connected firm. We compare the average corruption risk indicators of the 35 towns that awarded lighting contracts to Elios in the period in question with all other towns in our sample in Figure~\ref{fig:elios_dist}. Using a Mann-Whitney U-test, we find that towns involved in the scandal have significantly higher rates of corruption risk according to both measures (64\% vs 58\% $C_{csb}$ rate, U = 1385, p<.033; .30 vs .28 average CRI, U = 1397, p=.037). 

\subsection*{Measuring social capital}
iWiW was popular online social network operating in Hungary from 2004 to 2013. At its peak it boasted over 3.5 million active users (out of a population of around 10 million) and was among the top 3 most visited sites in the country. After a period of sustained popularity, it finally collapsed in 2013 as competitors, including Facebook, conquered the market. The increasing tendency of users to leave led to cascades of churners, highlighting the networked nature of the site~\cite{torok2017cascading,lorincz2017collapse}. Geographic distance is a major predictor of the likelihood of friendship ties on iWiW, and connections between towns reflect historical administrative boundaries and geographical barriers~\cite{lengyel2015geographies}. We use data from iWiW and measures based on its network of friendships to quantify bonding and bridging social capital in Hungarian settlements.

The iWiW network consists of users as nodes and mutually acknowledged friendship ties between users as links. Data from iWiW includes information on each user's town, selected from a menu. We used an anonymized version of the data to insure privacy. We describe steps we took to clean the network and the distribution of use rates at the settlement level in the SI.

Despite valid concerns about the representativity of data taken from online social networks~\cite{Torok2016big}, studies indicate data from online social networks offer a useful picture of the social capital of their users~\cite{williams2006and,brooks2014assessing}. As adoption of online social networks increases, they become increasingly useful for the study of the social structures~\cite{dunbar2015structure}. In any case, we control for possible confounding factors including settlement average income, rate of iWiW use, and share of the population over 60 in our models.

\subsubsection*{Fragmentation}
Our first settlement-level network measure, \textit{fragmentation}, quantifies the extent to which people in the settlement form densely connected and well separated subgroups. We do not consider the links residents of a settlement have with other towns. Fragmentation measures a settlement's bonding social capital.

We measure fragmentation of the settlement's internal social network using a community detection method to identify groups of highly connected nodes. We use the Louvain algorithm~\cite{blondel2008fast}, a popular and efficient method leading to a partition of the network. We measure the quality of the partition, the tendency of edges to be within rather than between the detected groups, using modularity~\cite{newman2006modularity}. Given a social network of users in a settlement $S$ and a partition of the network's nodes into $K$ groups, the modularity $Q(S)$ of the partition of the network can be written as:

$$Q(S) = \sum_{k=1}^{K} \Big[ \dfrac{L^{w}_{k}}{L} -\Big(\dfrac{L_{k}}{L}\Big)^{2} \Big],$$
where $L$ is the total number of edges in the considered network, $L_{k}$ is the number of edges adjacent to members of group $k$, and $L_{k}^{w}$ is the number of edges within group $k$. 

As modularity is highly dependent on the size and density of the network~\cite{Fortunato2009resolution}, we scale each settlement's modularity score in order to make valid comparisons between the towns. Following Sah et al.~\cite{sah2017unraveling}, we divide each settlement's modularity score by the theoretical maximum modularity $Q_{max}(S)$ that the given partition could achieve, namely if all edges were within groups. 

$$ Q_{\rm max}(S) = \sum_{k=1}^{K} \Big[ \dfrac{L_{k}}{L} -\Big(\dfrac{L_{k}}{L}\Big)^{2} \Big].$$

\noindent We then define the \textit{fragmentation} $F_S$ of a settlement $S$ as the quotient
\begin{equation}
F_S=Q(S)/Q_{\rm max}(S).
\label{eq:fragmentation}
\end{equation}

\noindent Fragmentation measures the tendency of individuals to belong to distinct groups within a settlement. A fragmented settlement consists of tightly-knit groups that are weakly connected. Both the excess of connections within and the rarity of connections between groups in fragmented networks are relevant to our theoretical framing of the origins of corruption as they indicate excess bonding social capital. The high density of connections within a group facilitates the enforcement of reciprocity, while having few connections between groups fosters particularism.

\begin{figure*}
\centering
  \includegraphics[width=.75\textwidth]{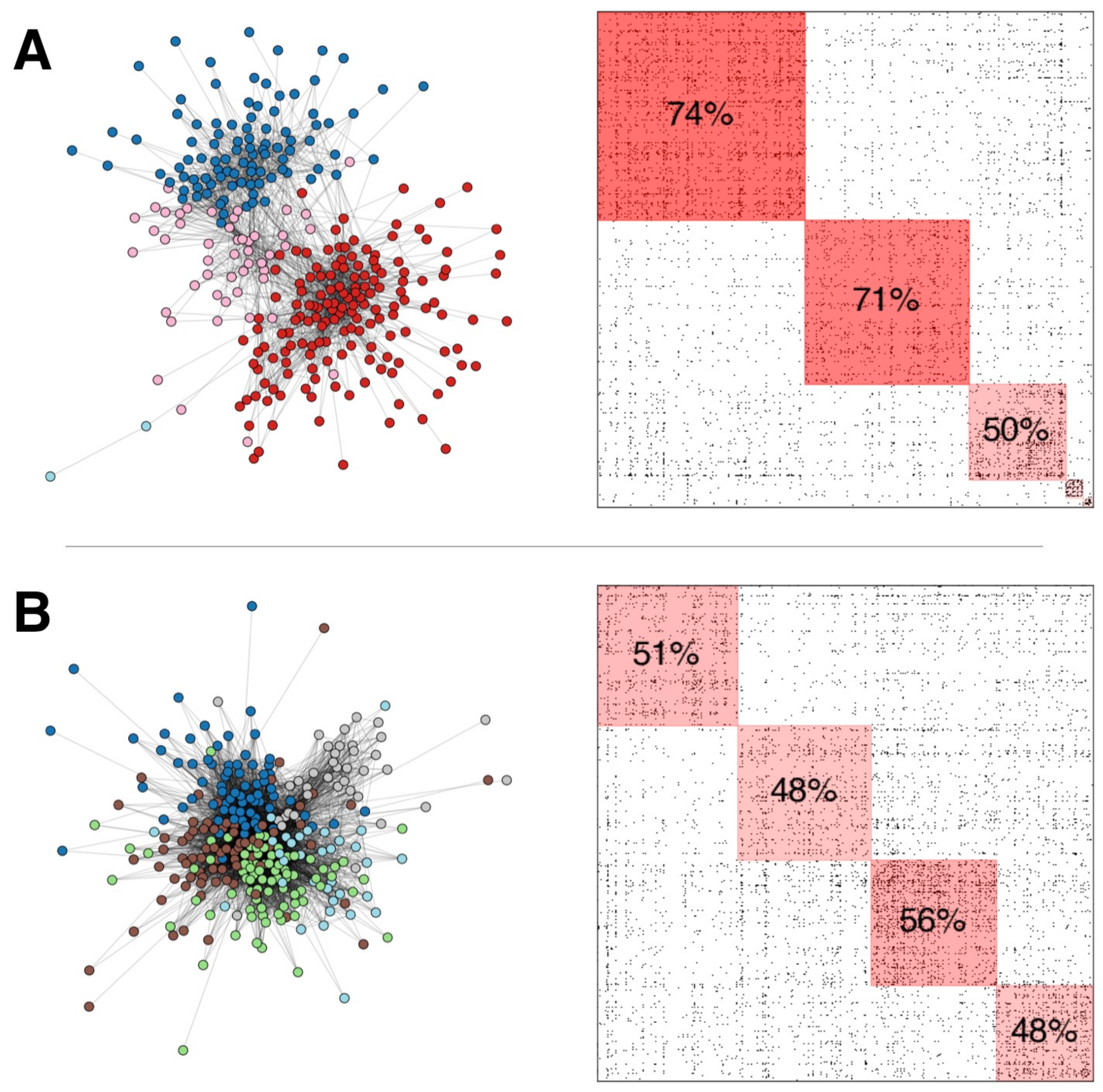}
  \caption{Sampled social networks and adjacency matrices of high (A) and low (B) fragmentation settlements. Node colors indicate membership in communities. In the adjacency matrices, percentages indicate the share edges staying within each community. In the fragmented settlement, communities have a significantly fewer connections with other communities.}
  \label{fig:town_nets}
\end{figure*}

To better understand the concept of fragmentation, we compare two settlements, one at the 90th percentile of fragmentation (town $A$) and the other at the 10th percentile (town $B$). The two towns have populations of roughly 10,000 and have iWiW user rates between 30 and 35\%. We randomly sample 300 users from both social networks for the sake of visualization and plot their connections in Figure~\ref{fig:town_nets}. Town $A$ is clearly more fragmented than town $B$. We also show the full adjacency matrices of the networks of these settlements, grouping nodes by their detected communities into blocks on the diagonal shaded in red. We label each community by the share of its edges staying within the community. The fragmented settlement has a clear over-representation of within-community edges.

\subsubsection*{Diversity}
Past research on online social networks noted that users connect with a variety of groups from their life-course~\cite{brooks2014assessing}. For example a user may connect with her schoolmates, university classmates, coworkers, family, and friends from environments including social clubs, sports teams, or religious communities. We measure the diversity of an individual user's network by the (lack of) overlap between these groups. In this case, we do not restrict our attention to edges between users from the same town.

\begin{figure*}[t]
\centering
  \includegraphics[width=.9\textwidth]{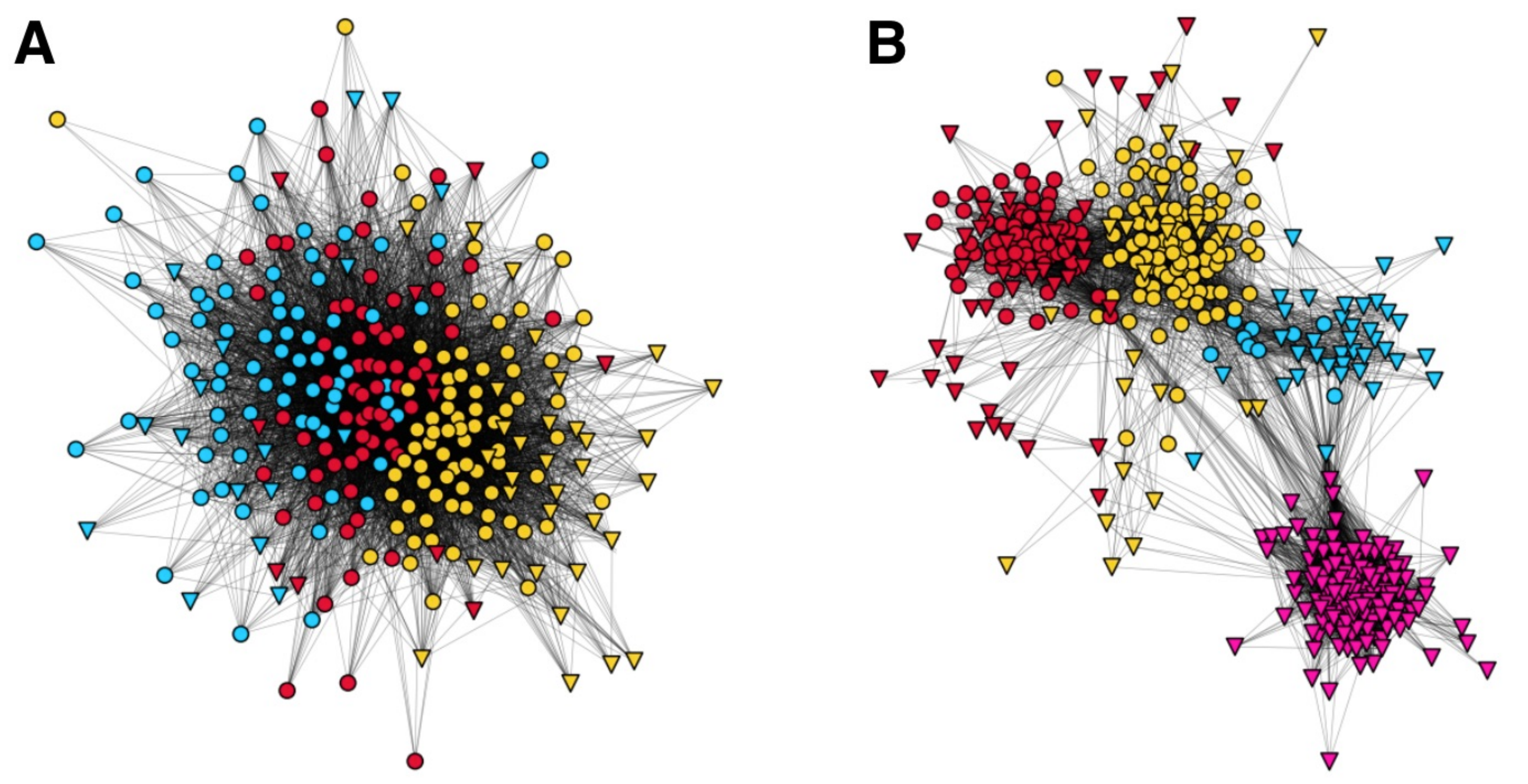}
  \caption{Ego networks with low (A) and high (B) diversity, respectively. Colors indicate membership in detected communities in the ego network. Circles denote users from the same settlement as the ego, while triangles mark users from elsewhere. The high diversity user's network has clusters of alters mostly from different towns.}
  \label{fig:ego_nets}
\end{figure*}

Following Brooks et al.~\cite{brooks2014assessing}, we consider the connections amongst the friends of a focal user or ego, i.e. the ego network without the ego. We then detect communities in the resulting network using the Louvain algorithm~\cite{blondel2008fast}. We can assume that members of a community of alters share some common context. We measure the separation of these communities of alters using modularity. Low modularity indicates that a user's connections tend to know each other, and that the user's different spheres of life involve the same people. High modularity indicates that the ego has a bridging role between weakly connected communities, and so we refer to such users as having high diversity in their social networks. We show examples of low and high diversity users with networks of similar sizes in Figure~\ref{fig:ego_nets}.

We aggregate this user level measure to a measure of settlement diversity $D_{S}$ by averaging each user's modularity score:

\begin{equation*}
D_S={1\over{|S|}}\sum_{i \in S}Q(\{\textrm{alters}_i\}),
\label{eq:diversity}
\end{equation*}

\noindent where $|S|$ is the number of nodes in the settlement $S$ and $\{\textrm{alters}_i\}$ is the subgraph of the alters of node $i\in S$. This measure captures the typical diversity of social perspectives that the members of the town access. At the settlement level this measure captures bridging social capital.

Settlement diversity is positively correlated with share of the population graduating from high school ($\rho\approx0.62$) and average income ($\rho \approx0.55$). Fragmentation and diversity are positively correlated ($\rho \approx 0.46$), which is not surprising given that both are calculated using network modularity. However, the ego-focus and, more importantly, the inclusion of inter-settlement edges of the diversity measure distinguish it from fragmentation, see the SI. Despite this correlation, we observe that they predict different corruption outcomes.

\subsection*{Models}
The primary aim of our paper is to relate bonding and bridging social capital in settlements to corruption risk in their public contracts. Our hypothesis H1 is related to excess bonding social capital, measured by fragmentation while hypothesis H2 refers surplus in bridging social capital, measured by diversity. We predict average contract corruption risk at the settlement level using Ordinary Least Squares (OLS) multiple regressions of the following form:

$$C_{S} = \beta_{1} * F_{S} + \beta_{2}*D_{S} + X_{S}*\theta + \epsilon_{S}$$

Where $C_{S}$ is one of two corruption risk indicators, averaged at the settlement level, $F_{S}$ is the settlement's fragmentation, $D_{S}$ is the settlement's diversity, $X_{S}$ is a matrix of control variables defined below, and $\epsilon_{S}$ is an error term. The $\beta$-s are scalar and $\theta$ a vector of  unknown parameters.

We include a variety of control variables in our regressions. Past research has found significant relationships between wealth, education, employment and corruption~\cite{mungiu2013controlling} so we control for settlement average income, its share of high school graduates, the presence of a university in the town, and its unemployment and inactivity rates. As demographic features of settlements may influence the measured social network we include total population, rate of iWiW use, and share of the population over 60 in our models~\cite{pfeil2009age}. We also control for the settlement's mayor's average victory margin in the 2002, 2006, 2010 elections as a proxy for the level of political competition in the town~\cite{broms2017procurement}. Finally, we include a geographic feature of the settlements: the minimum travel distance in minutes from the capital, Budapest. For the sake of comparison we fit a baseline models including only the control terms.

\section*{Results}
We summarize our findings in Table~\ref{tab:ols_regs}. We see that there is a significant relationship between social network structure and both dependent variables measuring corruption. More fragmentation consistently predicts more corruption, while more diversity consistently predicts less corruption. In both cases adding the network features significantly improves the adjusted R$^{2}$ of the model. Moreover, comparing the coefficients, we see that the social network features have effect sizes comparable to that of any social, political, or economic control. We present the full models in the SI, including the intermediate models containing only one network feature. All models pass a variance inflation factor (VIF) test for feature collinearity, see the SI.

\begin{table*}
\ra{1.2}
\setlength\tabcolsep{2.5pt}
\begin{tabular}{@{\extracolsep{1cm}}lp{1.6cm}p{1.6cm}p{1.6cm}p{1.6cm}} 
\toprule
\setlength\tabcolsep{2.5pt}
Dependent variable: & \multicolumn{2}{c}{\% Closed or single bid.} & \multicolumn{2}{c}{Average CRI} \\ 
\\[-1.8ex] & \multicolumn{1}{c}{(1)} & \multicolumn{1}{c}{(2)} &\multicolumn{1}{c}{(3)} & \multicolumn{1}{c}{(4)} \\ 
\cmidrule{2-3} \cmidrule{4-5}
 \textbf{\textit{Fragmentation}}  &  & \textbf{0.263$^{***}$} &  & \textbf{0.207$^{**}$} \\ 
  (Bonding social capital) &  & (0.097) &  & (0.092) \\ [.3cm]
 \textbf{\textit{Diversity}} &   & \textbf{$-$0.553$^{***}$} &  & \textbf{$-$0.551$^{***}$} \\ 
 (Bridging social capital) &    & (0.176) &  & (0.168) \\ [.3cm]
 Controls & Yes & Yes & Yes & Yes\\ [.15cm]
 Constant & 1.245 & 1.206 & 2.779 & 2.790\\ [.15cm]
Observations & 169 & 169 & 169 & 169 \\ 
Adjusted R$^{2}$ & 0.163 & \textbf{0.230} & 0.183 & \textbf{0.243} \\ 
F Statistic &  3.967$^{***}$ & 4.859$^{***}$ & 4.419$^{***}$  & 5.142$^{***}$  \\ 
\bottomrule
\end{tabular} 
  \caption{Town-level regression results predicting two corruption risk indicators. For both dependent variables, the first columns (1) and (3) correspond to the base model, predicting corruption risk using only control variables, and the second columns (2) and (4) show results, when the social network features are included. Note that all features are standardized with mean 0 and standard deviation 1. Full model tables are presented in the SI. Significance thresholds: $^{*}$p$<$0.1; $^{**}$p$<$0.05; $^{***}$p$<$0.01.} 
  \label{tab:ols_regs}
\end{table*}

\begin{figure*}[t]
\centering
 \includegraphics[width=.9\textwidth]{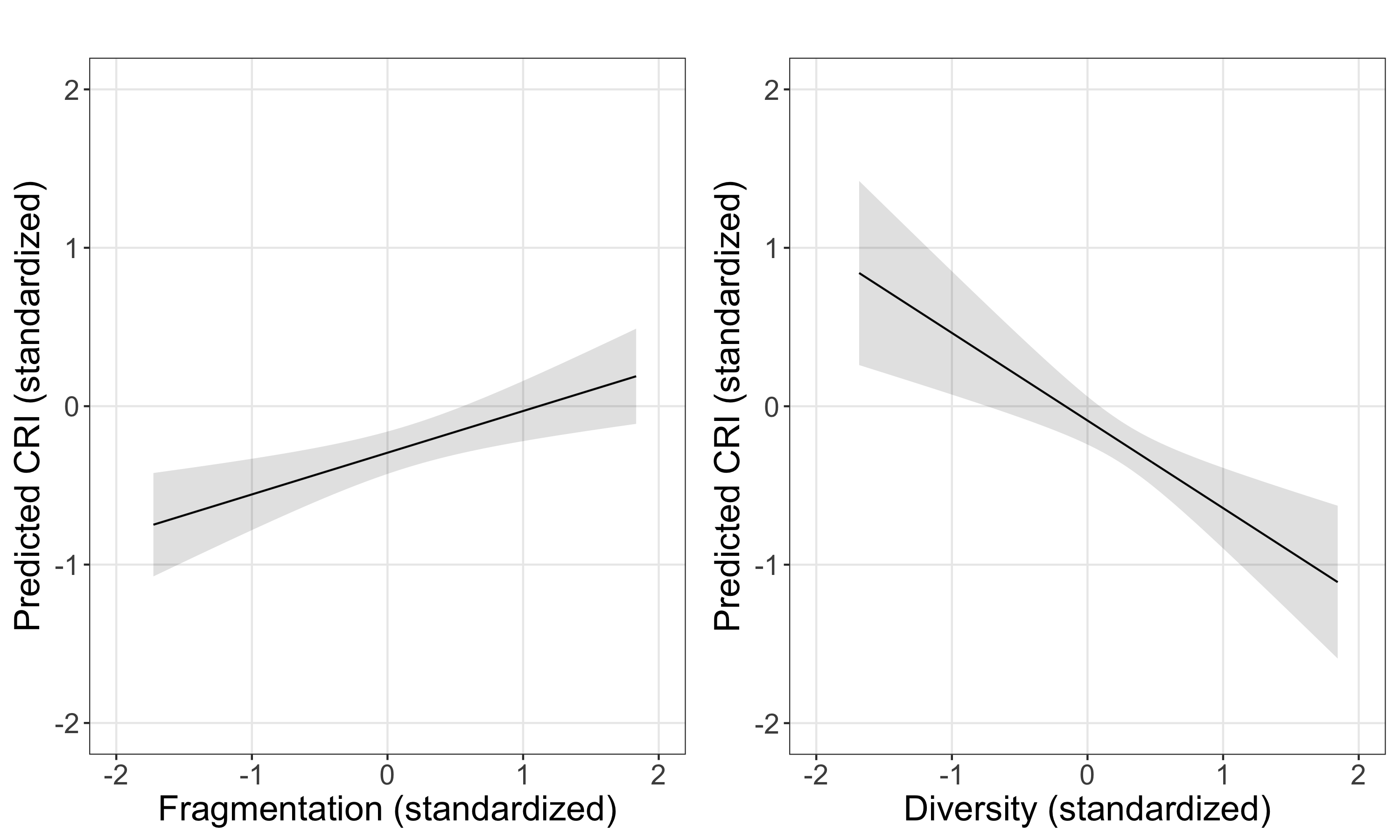}
 \caption{Plots of marginal effects of the key social capital variables and their predicted impact on a settlement's rate of closed procedure or single bidder contract awards; shaded regions represent 90\% confidence intervals are indicated. As the variables are standardized, unit changes on either axis can be interpreted as standard deviation changes. Fragmentation (on the left), quantifying excess bonding social capital in a community, predicts higher corruption risk, while diversity (on the right) predicts lower corruption risk.}
  \label{fig:olsmargins}
\end{figure*}

We visualize the effects of our network variables in Figure~\ref{fig:olsmargins}. We plot model predicted rate of closed procedure or single bid contract awards ($C_{csb}$) including 90\% confidence intervals for varying levels of fragmentation and average ego diversity. As the variables are standardized, the units can be interpreted as standard deviations from the mean (at 0). We observe that, all else equal, our model predicts that going from one standard deviation below average fragmentation to one standard deviation above average, increases $C_{csb}$ by about one half of a standard deviation. Diversity has a stronger effect in the other direction: the same change (from one standard deviation below average to one above average) induces a full standard deviation decrease in the corruption indicator. The effect of the network features on $C_{CRI}$ is similar. In the SI, we present an ANOVA feature importance test that indicates the significance of both network-based features.

\section*{Discussion}

In this paper we used data from an online social network and a collection of public procurement contracts to relate the social capital of Hungarian settlements to the corruption in its local government. To our knowledge, this paper is the first to study social aspects of corruption using large-scale social network data.

We introduced measures to quantify excess bonding and bridging social capital at the settlement level from online social network data. We found that settlements with high bonding social capital tend to award contracts with higher corruption risk. We also found that settlements with high bridging social capital, tend to award lower corruption risk contracts. Social capital measures add substantive predictive power to models of corruption outcomes, above baseline models controlling for other socio-economic factors such as average income, education, political competition, and demography.

We recognize several limitations to our approach. An inherent challenge in the research of corruption is that proven cases are rare, and so our measures can only track risk or suspicion of corruption. Moreover, we assume that steering contracts to certain firms by bureaucrats indicates corruption - but it may happen that bureaucrats make socially optimal decisions using their local knowledge of markets and discretion. iWiW is not a full map of social relations in Hungary and its users do not make up a representative sample of the population. Finally, we do not claim to have found a causal link between social capital and corruption risk. Besides the potential of omitted variable bias, corruption can also influence social capital~\cite{richey2010impact}.

Despite these limitations we believe that our findings are valuable. Above all, our novel, data-based, town-level approach provides new evidence for the old hypothesis that corruption is a structural phenomenon. This explains why appointing an ombudsman in a corrupt place rarely improves corruption outcomes~\cite{mungiu2013controlling} and why anti-corruption laws can backfire if they conflict with prevailing social norms~\cite{acemoglu2017social}. 

That is not to say that fighting corruption is futile. Rather we believe our findings suggest that top-down efforts are unlikely to work unless they impact social capital or other significant covariates of our model like political competition. Our conclusions hint at potential mechanisms which sustain corruption. Factors such as racial segregation or economic inequality which may drive fragmentation are ideal targets for policy interventions~\cite{rothstein2005all}. 
\subsection*{Author contributions statement}

J.W. and J.K. conceived of the presented idea. J.W. and B.L. collected data. J.W. and T.Y. developed the methods used. J.W. analyzed the data. All authors contributed to writing the manuscript.

\subsection*{Additional information}

The authors declare no conflict of interest.
\bibliographystyle{abbrvnat}
\bibliography{iwiw_corr_bib}

\newpage
\section{Supplementary Information}

\subsection{Description of iWiW data}
\label{SI:iwiwappendix}
\begin{figure}[b]
\centering
  \includegraphics[width=\textwidth]{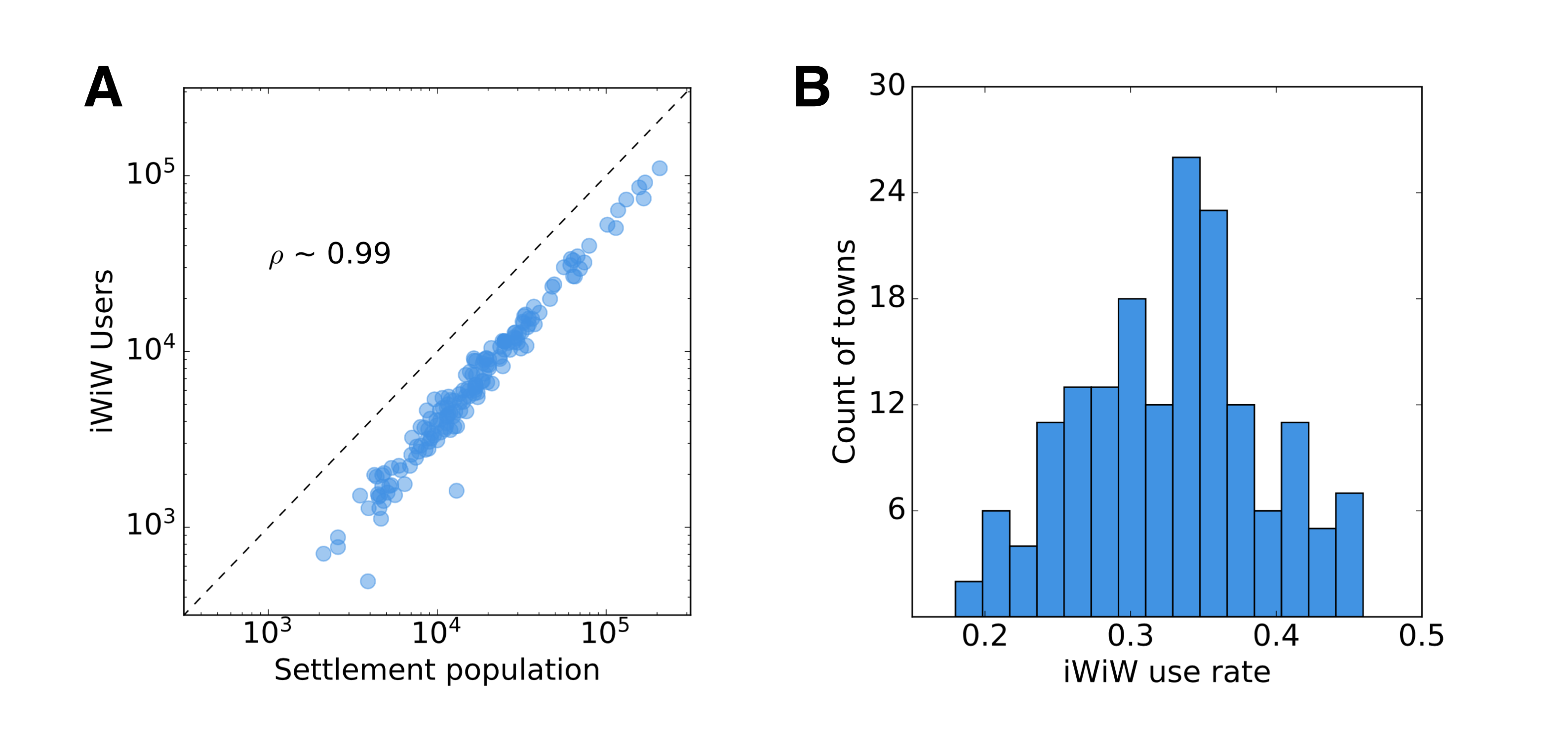}
  \caption{A) Settlement population and number of iWiW users plotted on a log-log scale. B) iWiW use rate by town.}
  \label{fig:iwiw_stats}
\end{figure}

In line with previous work on iWiW we filtered the data used in our analysis. Out of roughly 4.5 million user accounts, we dropped the roughly 500,000 accounts with location outside of Hungary. We also dropped the 193 users with more than 10,000 connections, arguing that such a large number of connections cannot represent social ties. In Figure~\ref{fig:iwiw_stats} we show the relationship between settlement population and the number of iWiW users listing their location in the settlement, and the share of the population registered to iWiW.

\subsection{Relationship between fragmentation and diversity}
\label{SI:fragdiv}
Fragmentation and diversity, our measures of bonding and bridging social capital respectively, are positively and significantly correlated ($\rho \approx0.46$). Though fragmentation considers only edges within the settlement and ego diversity includes external edges, both variables measure modularity in the network. However, according to our hypotheses, they are expected to capture different kinds of socialization. We found that despite their positive correlation these features have opposite relationships with our corruption risk measures: high fragmentation is positively and high diversity is negatively correlated with corruption risk. To test whether inter-settlement edges or the ego focus of diversity does more to distinguish the measure from fragmentation we recalculated the diversity considering only edges within the settlement. This alternative ``internal'' diversity measure is weakly correlated ($\rho \approx 0.28$) with fragmentation, and strongly correlated with diversity ($\rho \approx 0.72$). This suggests that both the connections to other towns and the ego-focus of the diversity measure distinguish fragmented towns from diverse ones.

\subsection{Model covariates and controls}
\label{appendix:descriptives}

In this appendix section we present the settlement-level variables used as controls in our models. We also report their summary statistics. Note that in our models, we scale all features to have mean 0 and standard deviation 1. 

\begin{itemize}
\item \textit{Average income per capita}: Wealthier places tend to be less corrupt~\cite{mungiu2013controlling} as competition for limited resources is expected to create greater incentive to cheat.
\item \textit{Population (log)}: Larger cities may have different contracting needs, different political and social norms, and different network characteristics.
\item \textit{Number of contracts awarded (log)}: Settlements contracting more frequently may be more experienced and may follow better practices. As larger groups of people are involved in contracting, corruption may become more difficult.
\item \textit{Rate of iWiW use}: The rate of iWiW use both proxies for the economic development of the settlement and controls for differences in observed social network structure resulting from differences in access to the web.
\item \textit{Average mayoral victory margin}: Measured across three elections (2002, 2006, 2010), this variable proxies for the lack of political competition in the settlement. The absence of political competition has been shown to correlate with corruption~\cite{broms2017procurement}.
\item \textit{Share of population with at least a high school diploma}: Education is typically correlated with better control of corruption~\cite{rothstein2005all}.
\item \textit{Share of working-age population inactive and unemployment rate}: Counting the long-term and short-term unemployed respectively, these variables quantify economic stagnation. The economic hardship connected with high unemployment is conjectured to worsen political corruption~\cite{sung2004democracy}.
\item \textit{The minimum travel distance to Budapest, the capital city}: This variable captures the physical isolation of the settlement from the main economic, political, and social hub of the country. Past research has shown that geographic isolation reduces accountability and increases corruption~\cite{campante2014isolated}. 
\item \textit{Share of population over 60 years old}: This variable controls for the over-representation of the elderly. The elderly are underrepresented on online social networks and tend to use these platforms differently than younger users~\cite{pfeil2009age}.
\item \textit{Whether the settlement has a university}: This variable controls for the presence of a place of higher education in the settlement, including local branches of universities headquartered elsewhere. this which inflates the number of young people, hence likely iWiW users in the town.
\end{itemize}

\begin{table} \centering 
\begin{tabular}{@{\extracolsep{5pt}}lccccc} 
\\[-1.8ex]\hline 
\hline \\[-1.8ex] 
Statistic & \multicolumn{1}{c}{N} & \multicolumn{1}{c}{Mean} & \multicolumn{1}{c}{St. Dev.} & \multicolumn{1}{c}{Min} & \multicolumn{1}{c}{Max} \\ 
\hline \\[-1.8ex] 
Closed procedure or single bidder & 169 & 0.59 & 0.15 & 0.21 & 0.92 \\ 
Average CRI & 169 & 0.28 & 0.04 & 0.16 & 0.40 \\ 
Fragmentation & 169 & 0.32 & 0.04 & 0.16 & 0.46 \\ 
Avg. ego diversity & 169 & 0.35 & 0.07 & 0.20 & 0.51 \\ 
Income per capita (thousands HUF) & 169 & 823.57 & 189.93 & 488.44 & 1,516.55 \\ 
N contracts (log) & 169 & 4.52 & 0.69 & 3.69 & 6.42 \\ 
Population (log) & 169 & 9.72 & 0.89 & 7.66 & 12.24 \\ 
Rate iWiW use & 169 & 0.33 & 0.06 & 0.18 & 0.46 \\ 
Average mayoral victory margin & 169 & 0.15 & 0.14 & 0.00 & 0.64 \\ 
\% high school graduates & 169 & 47.23 & 10.22 & 25.70 & 76.80 \\ 
Distance to Budapest (minutes) & 169 & 114.00 & 54.34 & 22.55 & 228.57 \\ 
Share of population inactive & 169 & 0.30 & 0.04 & 0.20 & 0.40 \\ 
Unemployment Rate & 169 & 0.06 & 0.01 & 0.03 & 0.09 \\ 
Share of population 60+ & 169 & 0.24 & 0.03 & 0.15 & 0.39 \\ 
Has university & 169 & 0.25 & 0.44 & 0 & 1 \\ 
\hline \\[-1.8ex] 
\end{tabular} 
\caption{Descriptive statistics of key settlement-level variables and controls.}
\label{tab:desc}
\end{table}

\subsection{Model results, diagnostics, and feature importances}

We present the full model results in Table~\ref{full_ols_regs}. Note that all variables are standardized with mean 0 and standard deviation 1. This aids interpretation, for example: a one standard deviation increase in the settlement's mayor's average margin of victory increases corruption risk by roughly one quarter of a standard deviation. We also present models including only one of the two network measures in Table~\ref{stepwise}. The effect and significance of both features is preserved when the other is excluded.

The estimated coefficients of the control variables and their levels of statistical significance offer additional insight into the phenomenon of corruption risk. Wealthier settlements are in general less corrupt, though the effect is not significant for CRI. Rate of iWiW use is not related with corruption risk and this does not change when we include the social capital features. The average mayoral victory margin is a highly significant positive predictor of corruption risk. One potential explanation is that mayors, who do not face significant competition do not fear being voted out of office if they are corrupt. Similarly towns that are far from Budapest, which our models predict to be significantly more corrupt, may be insulated from investigation by the central authorities simply by being out of the spotlight.

\begin{table}
\begin{center}
\ra{1}
\setlength\tabcolsep{2.5pt}
\begin{tabular}{@{\extracolsep{1cm}}lp{1cm}p{1cm}p{1cm}p{1cm}} 
\toprule
\setlength\tabcolsep{2.5pt}
Dependent variable: & \multicolumn{2}{c}{\% Closed or single bid.} & \multicolumn{2}{c}{Average CRI} \\
\\[-1.8ex] & \multicolumn{1}{c}{(1)} & \multicolumn{1}{c}{(2)} &\multicolumn{1}{c}{(3)} & \multicolumn{1}{c}{(4)} \\ 
\cmidrule{2-3} \cmidrule{4-5}
 \textbf{\textit{Fragmentation}}  &  & \textbf{0.263$^{***}$} &  & \textbf{0.207$^{**}$} \\ 
  (Bonding social capital) &  & (0.097) &  & (0.092) \\ [.3cm]
 \textbf{\textit{Diversity}} &   & \textbf{$-$0.553$^{***}$} &  & \textbf{$-$0.551$^{***}$} \\ 
 (Bridging social capital) &    & (0.176) &  & (0.168) \\ [.3cm]
 Income/capita  & $-$0.262 & $-$0.277$^{*}$ & $-$0.075& $-$0.096 \\ 
  &  (0.169) & (0.162) & (0.161) & (0.155) \\ [.3cm]
 N contracts (log)  & $-$0.313$^{*}$ & $-$0.314$^{*}$ & $-$0.685$^{***}$ & $-$0.697$^{***}$ \\ 
  &  (0.171) & (0.165) & (0.162) & (0.158) \\ [.3cm]
 Population (log) & $-$0.180 & 0.020 & 0.118 & 0.335$^{**}$ \\ 
   & (0.143) & (0.166) & (0.136) & (0.159) \\ [.3cm]
 Rate iWiW use  & 0.045 & 0.037 & 0.122 & 0.107 \\ 
   & (0.137) & (0.132) & (0.130) & (0.126) \\ [.3cm]
 Mayor victory margin & 0.278$^{***}$ & 0.255$^{***}$ & 0.303$^{***}$ & 0.281$^{***}$ \\ 
  & (0.089) & (0.086) & (0.085) & (0.082) \\[.3cm]
 \% high school grads & 0.166 & 0.374$^{*}$ & $-$0.176 & 0.040 \\ 
  & (0.190) & (0.199) & (0.181) & (0.190) \\ [.3cm]
 Distance to Budapest  & $-$0.021 & $-$0.198$^{*}$ & 0.061 & $-$0.112 \\ 
   & (0.104) & (0.112) & (0.099) & (0.107) \\[.3cm]
 Share of pop. inactive &  $-$0.797$^{***}$ & $-$0.805$^{***}$ & $-$0.716$^{***}$ & $-$0.754$^{***}$ \\
  &  (0.229) & (0.229) & (0.218) & (0.219) \\ [.3cm]
 Unemployment Rate & 0.239$^{**}$ & 0.262$^{**}$ & 0.299$^{***}$ & 0.320$^{***}$ \\ 
 & (0.118) & (0.113) & (0.112) & (0.108) \\ [.3cm]
 \% population 60+ & 0.501$^{***}$ & 0.491$^{***}$ & 0.500$^{***}$ & 0.503$^{***}$ \\ 
  & (0.163) & (0.158) & (0.155) & (0.151) \\ [.3cm]
 Has university  & 0.351 & 0.294 & 0.431$^{**}$ & 0.352$^{*}$ \\ 
  & (0.220) & (0.221) & (0.210) & (0.211) \\ [.3cm]
 Constant  & 1.245$^{*}$ & 1.206$^{*}$ & 2.779$^{***}$ & 2.790$^{***}$ \\ 
   & (0.725) & (0.702) & (0.689) & (0.671) \\ [.15cm]
\hline
Observations & 169 & 169 & 169 & 169 \\ 
Adjusted R$^{2}$ & 0.163 & \textbf{0.230} & 0.183 & \textbf{0.243} \\ 
F Statistic &  3.967$^{***}$ & 4.859$^{***}$ & 4.419$^{***}$  & 5.142$^{***}$  \\ 
\bottomrule
\end{tabular} 
  \caption{Town-level regression results predicting two corruption risk indicators. For both dependent variables, the first columns (1) and (3) correspond to the base model, predicting corruption risk using only control variables, and the second columns (2) and (4) show results, when the social network features are included. Note that all features are standardized with mean 0 and standard deviation 1. Significance thresholds: $^{*}$p$<$0.1; $^{**}$p$<$0.05; $^{***}$p$<$0.01.} 
  \label{full_ols_regs}
  \end{center}
\end{table}

\begin{table}
\begin{center}
\ra{1}
\setlength\tabcolsep{2.5pt}
\begin{tabular}{@{\extracolsep{1cm}}lp{1cm}p{1cm}p{1cm}p{1cm}} 
\toprule
\setlength\tabcolsep{2.5pt}
Dependent variable: & \multicolumn{4}{c}{\% Closed or single bid.} \\
\\[-1.8ex] & \multicolumn{1}{c}{(1)} & \multicolumn{1}{c}{(2)} &\multicolumn{1}{c}{(3)} & \multicolumn{1}{c}{(4)} \\ 
\cmidrule{2-5}
 \textbf{\textit{Fragmentation}} &  &  & 0.233$^{**}$ & 0.263$^{***}$ \\ 
 (Bonding social capital) &  &  & (0.099) & (0.097) \\ 
  & & & & \\ 
 \textbf{\textit{Diversity}} &  & $-$0.505$^{***}$ &  & $-$0.553$^{***}$ \\ 
 (Bridging social capital) &  & (0.179) &  & (0.176) \\ 
  & & & & \\ 
 Income/capita & $-$0.262 & $-$0.295$^{*}$ & $-$0.243 & $-$0.277$^{*}$ \\ 
  & (0.169) & (0.166) & (0.167) & (0.162) \\ 
  & & & & \\ 
 N contracts (log) & $-$0.313$^{*}$ & $-$0.359$^{**}$ & $-$0.269 & $-$0.314$^{*}$ \\ 
  & (0.171) & (0.168) & (0.169) & (0.165) \\ 
  & & & & \\ 
 Population (log) & $-$0.180 & 0.083 & $-$0.257$^{*}$ & 0.020 \\ 
  & (0.143) & (0.168) & (0.144) & (0.166) \\ 
  & & & & \\ 
 Rate iWiW use & 0.045 & 0.009 & 0.073 & 0.037 \\ 
  & (0.137) & (0.134) & (0.135) & (0.132) \\ 
  & & & & \\ 
 Mayor victory margin  & 0.278$^{***}$ & 0.259$^{***}$ & 0.276$^{***}$ & 0.255$^{***}$ \\ 
  & (0.089) & (0.087) & (0.088) & (0.086) \\ 
  & & & & \\ 
 \% high school grads & 0.166 & 0.397$^{*}$ & 0.126 & 0.374$^{*}$ \\ 
  & (0.190) & (0.203) & (0.188) & (0.199) \\ 
  & & & & \\ 
 Distance to Budapest & $-$0.021 & $-$0.169 & $-$0.035 & $-$0.198$^{*}$ \\ 
  & (0.104) & (0.114) & (0.102) & (0.112) \\ 
  & & & & \\ 
 Share of pop. inactive & $-$0.797$^{***}$ & $-$0.931$^{***}$ & $-$0.675$^{***}$ & $-$0.805$^{***}$ \\ 
  & (0.229) & (0.229) & (0.232) & (0.229) \\ 
  & & & & \\ 
 Unemployment Rate & 0.239$^{**}$ & 0.253$^{**}$ & 0.247$^{**}$ & 0.262$^{**}$ \\ 
  & (0.118) & (0.115) & (0.116) & (0.113) \\ 
  & & & & \\ 
 \% population 60+ & 0.501$^{***}$ & 0.546$^{***}$ & 0.449$^{***}$ & 0.491$^{***}$ \\ 
  & (0.163) & (0.160) & (0.162) & (0.158) \\ 
  & & & & \\ 
 Has University & 0.351 & 0.198 & 0.449$^{**}$ & 0.294 \\ 
  & (0.220) & (0.222) & (0.221) & (0.221) \\ 
  & & & & \\ 
 Constant & 1.245$^{*}$ & 1.426$^{**}$ & 1.036 & 1.206$^{*}$ \\ 
  & (0.725) & (0.712) & (0.720) & (0.702) \\ 
  & & & & \\ 
\hline \\[-1.8ex] 
Observations & 169 & 169 & 169 & 169 \\ 
Adjusted R$^{2}$ & 0.163 & 0.198 & 0.186 & 0.230 \\ 
F Statistic & 3.967$^{***}$  & 4.460$^{***}$ & 4.207$^{***}$ & 4.859$^{***}$  \\ 
\bottomrule
\end{tabular} 
  \caption{Stepwise regressions. The effect and significance of the network features are preserved when including them only one at a time. $^{*}$p$<$0.1; $^{**}$p$<$0.05; $^{***}$p$<$0.01.} 
  \label{stepwise}
  \end{center}
\end{table}

\label{appendix:model_diag}
\begin{figure*}
\includegraphics[width=\textwidth]{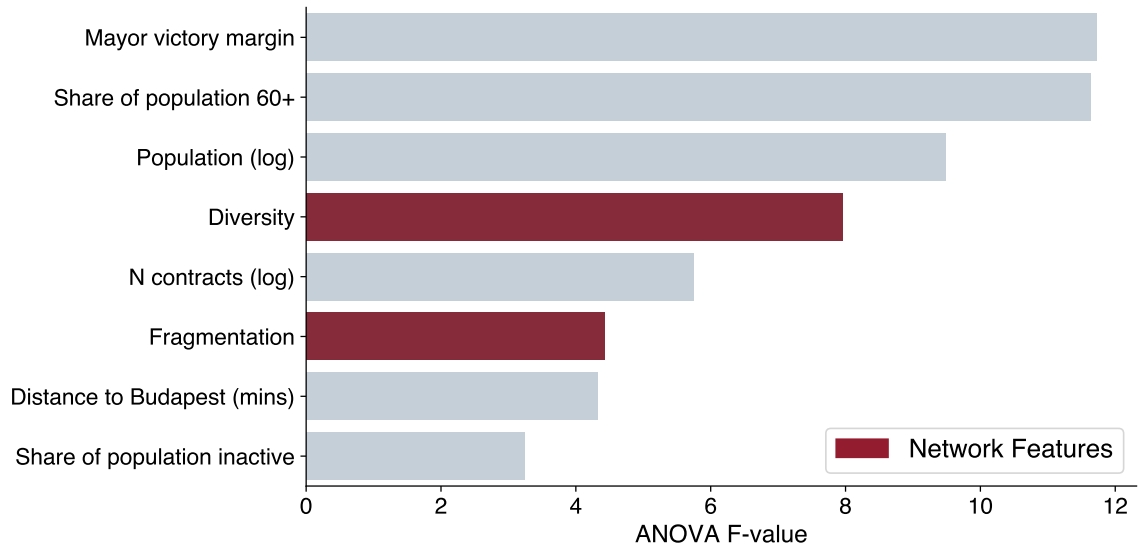}
\caption{Analysis of Variance F-test feature importances of OLS regression predicting average settlement CRI. We only include significant features, and highlight the network-based social capital measures.}
\label{fig:feat_imps}
\end{figure*}
One potential source of bias in the coefficient estimates of multiple regression models is collinearity among the predictors. We test for multi-collinearity for each predictor using a variance inflation factor (VIF) test, defined as the ratio of variance in the full model over the variance of the single-predictor model. We run this diagnostic for each predictor used in models (2) and (4) in the main text and report the results in Table~\ref{table:vif}. A popular rule of thumb is that VIF values under 10 denote acceptable levels of correlation between variables~\cite{hair1999analysis}.

\begin{table}
\begin{center}
\ra{1.1}
\begin{tabular}{ l  l   }\toprule
Predictor & VIF   \\\midrule
\textbf{\textit{Fragmentation}} &1.407\\ 
\textbf{\textit{Diversity}} & 6.337 \\
Income/capita & 5.430 \\ 
N contracts (log)& 3.045  \\
Population (log) & 5.892 \\ 
Rate iWiW use& 2.885 \\
Mayor victory margin & 1.040 \\ 
\% high school grads & 7.106  \\
Share of pop. inactive & 9.899 \\ 
Unemployment Rate & 2.360  \\
Distance to Budapest & 3.068 \\ 
\% population 60+ & 5.442  \\
Has university & 2.192  \\
\bottomrule
\label{table:vif}
\end{tabular}
\caption{VIF scores for model predictors.}
\end{center}
\end{table}

We show the relative variable importances of Model (6) (column 6 in Table~\ref{tab:ols_regs}), the fully specific model predicting average CRI, using an Analysis of Variance F-test in Figure~\ref{fig:feat_imps}. We include only terms with a significant ANOVA F-test. Though other features have stronger predictive power, the social network features are more useful in predicting corruption risk than economic variables like unemployment, inactivity, and average income.

\end{document}